\documentstyle[psfig]{mn}

\title{Dynamical friction for compound bodies}

\author[R. Dom\'{\i}nguez-Tenreiro \& M.A. G\'omez-Flechoso]{R. 
Dom\'{\i}nguez-Tenreiro \& M.A. G\'omez-Flechoso\thanks{Present
address: Observatoire de Gen\`eve, Ch. des Maillettes 51, Ch-1290 Sauverny
(Switzerland)}\\
Dept. F\'{\i}sica Te\'orica, C-XI. Univ. Aut\'onoma de Madrid, 
E-28049 Madrid}

\date{Accepted .....
      Received .....;
      in original form 1997 May 14}

\pagerange{\pageref{firstpage}--\pageref{lastpage}}

\begin{document}
\maketitle

\label{firstpage}

\begin{abstract}

In the framework of the fluctuation-dissipation approach to
dynamical friction, we derive an expression giving the orbital
energy exchange experienced by a compound body as it moves 
interacting with a non homogeneous discrete background. The
body is assumed to be composed of particles endowed with
a velocity spectrum and with a non homogeneous spatial 
distribution. The Chandrasekhar formula is recovered in the 
limit of a point-like satellite with zero velocity dispersion
and infinite temperature moving through an homogeneous infinite
medium. In this same limit, but dropping the zero satellite  velocity
dispersion ($\sigma_S$) condition, the orbital energy loss is 
found to  be smaller than in the $\sigma_S=0$ case by a factor of
up to an order of magnitude in some situations.

\end{abstract}


\section{Introduction}

A satellite moving through a field of gravitating particles experiences a
dissipative frictional force known as dynamical friction. It can be understood
in terms of the satellite wake, that exerts a drag force on the satellite
itself (Kalnajs 1972; Binney \& Tremaine 1987; Weinberg 1989; Bekenstein 
\& Zamir 1991). It can also been understood in terms of the underlying
basic physics as the friction resulting from the fluctuating gravitating
forces acting on the satellite as a consequence of the non-continuos
character of the particle system. The fact that fluctuating forces cause 
dissipation is quite a general scheme in Physics. It is at the basis of
physical phenomena such as electric resistance in conductors or viscous
friction in liquids (Reif 1965).

Dynamical friction has important consequences in the evolution of astronomical
systems, mainly because it causes a decay of orbiting bodies, so that 
merger timescales and dissipation rates by dynamical friction are closely
related. The merger scenario is at the basis of a great deal of processes in 
Astronomy. Not only is it the framework for the general galaxy formation picture
in hierarchical cosmological models, 
but also for more particular
aspects of the evolution of a number of astronomical systems, such as
galactic nuclei, cD galaxies in rich galaxy clusters, compact galaxy groups and
so on.

A dynamical friction formula was first obtained, in a kinematical approach, by 
Chandrasekhar (1943). He has calculated the rate of momentum exchange 
between the test and field particles as the result of a sum of uncorrelated
two-body encounters, obtaining the expression:

\begin{equation}
M {d \mbox{\boldmath $v$} \over dt} = - {4 \pi G^{2} M^{2} \over v^{3}}  \ln \Lambda
\, \rho (< v)\,  \mbox{\boldmath $v$} ,
\label{Chf}
\end{equation}

\noindent
where $M$ and $\mbox{\boldmath $v$}$ are the test particle mass and speed, respectively,
$\rho (< v)$ is the density of background particles with velocity less than
$v$, $G$ is the gravitation constant and $\Lambda = {p_{max} \over p_{min}}$,
with $p_{max}$ and $p_{min}$ the maximum and minimum impact parameters
contributing to the drag. Keplerian orbits for both the test and field
particles and an infinite and homogeneous background have been assumed in the
derivation of eq. (\ref{Chf})

Chandrasekhar's formula is widely employed to quantify dynamical friction 
in a variety of situations, even if in most astronomical problems the 
background is neither infinite nor homogeneous. This formula is known to
give the correct order of magnitude, but it suffers from several drawbacks,  
that
arise from the very physical assumptions made in its derivation. Furthermore,
it cannot describe some situations, as for example the drag experienced by
a satellite placed outside the edge of a finite gravitating system. In fact,
the satellite would be decelerated
on physical grounds, because it causes a perturbation to the
system. However, according to eq. (\ref{Chf}), the drag would vanish. For this
reason, other works on dynamical
friction followed Chandrasekhar's pioneering study, either from a numerical 
(Lin \& Tremaine 1983; White 1983; Bontekoe \& van Albada 1987; Zaritsky
\& White 1988) or an analytical point of view. Analytical descriptions have
the advantage that they help understanding the underlying physics. Several
 methods have been developed: Fokker-Planck equation based 
(Rosenblunth, Mc Donald \& Judd 1957; Binney \& Tremaine 1987)
 polarization cloud approach (Marochnik 1968; Kalnajs 1972; 
Binney \& Tremaine 1987;
Weinberg 1989; Bekenstein \& Zamir 1991), resonant particle interactions
(Lynden-Bell \& Kalnajs 1972; Tremaine \& Weinberg 1984; Palmer \& 
Papaloizou 1985; Weinberg 1986).

Berkenstein and Maoz (1992, hereafter BM92) and Maoz (1993, hereafter M93)
introduced a fluctuation-dissipation approach to dynamical friction. The
fluctuation-dissipation theorem (Kubo 1959) relates the friction coefficient 
to the time integral of the correlation function for the fluctuations
causing the friction. They showed that dynamical friction fits into this 
general scheme, which provides a powerful technique to study it. This 
approach is in fact a return to Chandrasekhar's original attempt to give 
a statistical description of dynamical friction. Other 
stochastic approaches to dynamical friction are those of 
Cohen (1975) and Kandrup (1980). M93 derived a formula for
the drag experienced by an object which travels in an arbitrary mass density 
field, assumed to be stationary and formed by particles much lighter than
the object.

A common feature of all the previous approaches is that they treat the satellite
as being rigid and without structure in the velocity space. This approximation 
can be  good enough in a number of astronomical situations, but in others,
these two 
ingredients could play a crucial role. As a first example of such a situation,
let us consider the dynamical evolution of compact groups of galaxies 
(e.g. Mamon 1993). In this case, the velocity dispersion of individual galaxies
is comparable to the velocity dispersion of the common halo that hosts them.
As a second example, we recall that in the problem of the 
interaction of two comparable
mass galaxies, the energy exchange due to dynamical friction cannot be
calculated in the previous frameworks, because  their presumably comparable 
velocity dispersions need to be taken into account.

In this paper we present a fluctuation-dissipation study of the orbital changes
experienced by  a non-rigid satellite, composed of gravitating  particles 
with a finite velocity dispersion, as it
interacts with a general background. An extension
of BM92 and M93 techniques has allowed us to calculate the rate of energy 
exchange between them as a result of fluctuations in the gravitational forces
of both the background and the satellite.

The paper is organized as follows: in Sect. 2, the physical formulation of the 
method and the general expressions giving the energy exchange rate 
are presented. In Sect. 3, we calculate the instantaneous energy variations
for general backgrounds at rest and with a Maxwellian velocity distribution.
Some particular limits are dealt with in Sect. 4. Finally, in Sect. 5,
we summarize and discuss our results. Two Appendices follow, where the 
results of the calculation of the correlation matrix and of an integral
needed are given.

\section{Physical formulation}

We will study the energy exchange between two self-gravitating equilibrium
systems called the background (B) and the satellite (S). The background
consists of $N_B$ equal mass particles, each of them with a mass $m_B$.
Its total mass and typical size are $M_B = m_B N_B$ and $R_B$, respectively.
These particles exhibit a velocity spectrum with dispersion $\sigma_B$ and
zero mean at $t = 0$ (which is equivalent to considering the origin
of the reference system placed at the halo center of mass
at $t=0$). The satellite in turn is  composed of $N_S$ equal 
mass particles of mass $m_S$, total mass $M_S$, typical size $R_S$, and a 
velocity distribution with dispersion $\sigma_S$ and mean equal to the center
of mass velocity of the satellite, $\mbox{\boldmath $v$}_{CMS}$.

As we are mainly interested in the effects that a non vanishing $\sigma_{S}$
would have on the evolution of the whole system, we can  consider that
$\sigma_S \sim \sigma_B \equiv \sigma$. The virial theorem tells us that
$\sigma \sim G N m R^{-1}$, so that $\sigma_S \sim \sigma_B$ implies that

\begin{equation}
N_S m_S R^{-1}_S \sim N_B m_B R^{-1}_B.
\label{sim}
\end{equation}

As both the satellite and the background are assumed to have a finite size, 
we can safely consider that

\begin{equation}
N_{\alpha}^{1/n} \gg \left( {R_{\beta} \over R_{\gamma}} \right)^m
\label{gg}
\end{equation}

\noindent
for whatever $\alpha, \beta, \gamma = B$ or $S$, and $n, m$ of order unity.
If $m_{\alpha} \gg m_{\tilde{\alpha}}$ ($\tilde{\alpha}$ is the particle
class contrary to $\alpha$), as could be the case, for example,
when the $\alpha$ and $\tilde{\alpha}$ systems are composed of stars and  
dark matter particles, respectively, then eqs. (\ref{sim}) and (\ref{gg})
imply the inequality $N_{\alpha} \ll N_{\tilde{\alpha}}$. But, in any case,
both $N_{\alpha}$ and $N_{\tilde{\alpha}}$ will be assumed to be very large.

\subsection{The fluctuating forces acting on a particle}

The fluctuating force, $\mbox{\boldmath $F$}_{i_S}(\mbox{\boldmath $r$}_{i_S},t)$, acting 
at time $t$ on a satellite 
particle $i_S$ placed at position $\mbox{\boldmath $r$}_{i_S}$ and caused by its
interactions with both background and satellite particles, can be written
as:

\begin{equation}
\mbox{\boldmath $F$}_{i_S}(\mbox{\boldmath $r$}_{i_S},t) = \mbox{\boldmath $F$}^{B}_{i_S}(\mbox{\boldmath $r$}_{i_S},t) +
				 \mbox{\boldmath $F$}^{S}_{i_S}(\mbox{\boldmath $r$}_{i_S},t)
\label{ffts}
\end{equation}

\noindent
where

\begin{eqnarray}
\mbox{\boldmath $F$}^{B}_{i_S}(\mbox{\boldmath $r$}_{i_S},t) &=& -G m_S m_B 
\mbox{\boldmath $\nabla$}
\left( \sum_{j_B} {1 \over \mid \mbox{\boldmath $r$}_{i_S} - \mbox{\boldmath $r$}_{j_B} \mid } \right. \nonumber \\
&-& \left. \int { d\mbox{\boldmath $r$} n_B(\mbox{\boldmath $r$}) \over 
\mid \mbox{\boldmath $r$}_{i_S} - \mbox{\boldmath $r$} \mid} \right )
\label{ffbs}
\end{eqnarray}

\noindent
and

\begin{eqnarray}
\mbox{\boldmath $F$}^{S}_{i_S}(\mbox{\boldmath $r$}_{i_S},t) &=& -G m_S^{2}  
\mbox{\boldmath $\nabla$}
\left( \sum_{j_S \neq i_S} {1 \over \mid \mbox{\boldmath $r$}_{i_S} - \mbox{\boldmath $r$}_{j_S} \mid } \right. \nonumber \\
&-& \left. {N_{S}-1 \over N_S} \int { d\mbox{\boldmath $r$} n_S(\mbox{\boldmath $r$}) \over
\mid \mbox{\boldmath $r$}_{i_S} - \mbox{\boldmath $r$} \mid} \right )
\label{ffss}
\end{eqnarray}

\noindent
are the fluctuating forces acting on $i_S$ caused by the background and
satellite, respectively. Each of these forces results from subtracting
to the total many-body or discrete force a smooth part derived from the mean
field potentials, $\Phi_B(\mbox{\boldmath $r$})$ and $\Phi_S(\mbox{\boldmath $r$})$, due to the
smooth densities, $n_B(\mbox{\boldmath $r$})$ and  $n_S(\mbox{\boldmath $r$})$, respectively. These
satisfy the relations:

\begin{equation}
n_{\alpha}(\mbox{\boldmath $r$}) = N_{\alpha} \int d\mbox{\boldmath $u$} f_{0}^{\alpha}(\mbox{\boldmath $r$}, \mbox{\boldmath $u$}),
\label{defn}
\end{equation}

\begin{equation}
\nabla ^2 \Phi_{\alpha}(\mbox{\boldmath $r$}) = 
4 \pi G m_{\alpha} n_{\alpha}(\mbox{\boldmath $r$})
\label{poisson}
\end{equation} 

\noindent
and 

\begin{equation}
\int d\mbox{\boldmath $r$} n_{\alpha}(\mbox{\boldmath $r$}) = N_{\alpha},
\label{norm}
\end{equation}

\noindent
where $\alpha=B,S$, 
$f_{0}^{\alpha}(\mbox{\boldmath $r$}, \mbox{\boldmath $u$})$ 
is the one-particle  distribution function for
background and satellite particles in the unperturbed state, and the 
${N_S-1 \over N_S}$ factor takes into account that the $i_S$ particle
does not interact with itself at time $t$. An exchange of the $B$ and $S$
labels in eq. (\ref{ffbs}) gives the expression for the fluctuating force
on the background particle $i_B$ caused by the satellite. Changing the $S$
labels into $B$ in eq. (\ref{ffss})
we get the force on $i_B$ caused by the background fluctuations. 
In a compact formulation, we can write the fluctuating force on a generic
class $\alpha$ particle, $i_{\alpha}$, due to class $\beta$ particles as:

\begin{eqnarray}
\mbox{\boldmath $F$}^{\beta}_{i_{\alpha}}(\mbox{\boldmath $r$}_{i_{\alpha}},t) &=& 
-G m_{\alpha} m_{\beta} 
\mbox{\boldmath $\nabla$}
\left( \sum_{j_{\beta} \neq i_{\alpha}} {1 \over 
\mid \mbox{\boldmath $r$}_{i_{\alpha}} - \mbox{\boldmath $r$}_{j_{\beta}} \mid } \right. \nonumber \\
&-& \left.{N_{\beta}^{'} \over N_{\beta}} \int { d\mbox{\boldmath $r$} n_{\beta}(\mbox{\boldmath $r$}) \over
\mid \mbox{\boldmath $r$}_{i_{\alpha}} - \mbox{\boldmath $r$} \mid} \right )
\label{ffgen}
\end{eqnarray}

\noindent
where now $N_{\beta}^{'} = N_{\beta} - \delta_{{\alpha}{\beta}}$ 
takes into
account the possibility that ${\alpha} = {\beta}$.

The equation of motion of either a background or a satellite particle at
time $t$ can be written in a compact notation as:

\begin{eqnarray}
\mbox{\boldmath $v$}_{i_{\alpha}} (t) &=& \mbox{\boldmath $v$}_{i_{\alpha}, 0} +
\sum_{\beta} \left[ {1 \over m_{\alpha}} \int_{0}^{t} 
\mbox{\boldmath $F$}^{\beta}_{i_{\alpha}}(\mbox{\boldmath $r$}_{i_{\alpha}},t') dt' \right. \nonumber \\
&-& \left. \int_{0}^{t} \mbox{\boldmath $\nabla$} \Phi_{\beta}(\mbox{\boldmath $r$}_{i_{\alpha}},t') 
dt'\right] ,
\label{eqmov}
\end{eqnarray}

\noindent
where $\mbox{\boldmath $v$}_{i_{\alpha}, 0}$ is the particle velocity at time $t =0$.

Stochastic forces are weak as compared with the smooth global forces. In fact,
the fluctuating force between a class $\alpha$ and a class $\beta$ particle
is of order (Kandrup 1980, Saslaw 1985)

\begin{equation}
\mid \mbox{\boldmath $F$}^{\beta}_{i_{\alpha}, fluc} \mid 
=  O \left( {G m_{\alpha} m_{\beta} n^{2/3}} \right) ,
\label{Ofluc}
\end{equation}

\noindent
where $n \sim N_{\alpha} R_{\alpha}^{-3} + (1 - \delta_{\alpha \beta})
N_{\beta} R_{\beta}^{-3}$ is the average particle number density for the whole 
system. The smooth density of class $\delta$ particles causes a gravitating
force on a class $\gamma$ particle that is of order:

\begin{equation}
\mid \mbox{\boldmath $F$}^{\delta}_{j_{\gamma}, smooth} \mid = 
 O \left( G m_{\gamma}
m_{\delta} N_{\delta} R_{\delta}^{-2} \right) .
\label{Osmooth}
\end{equation}

Eqs. (\ref{Ofluc}) and (\ref{Osmooth}) give, after some algebra:

\begin{eqnarray}
{\mid \mbox{\boldmath $F$}^{\beta}_{i_{\alpha}, fluc} \mid \over
\mid \mbox{\boldmath $F$}^{\delta}_{j_{\gamma}, smooth} \mid }\!\!&\!=
\!&\!\! O \left( \left[ \left(
{ m_{\beta} \over m_{\delta}} \right)^{1/2} 
 \left( { R_{\delta} \over R_{\beta}} \right)^{2} \right. \right. 
\nonumber \\
\!&\!+\!&\!\!\! \left. \left. (1 - \delta_{\alpha \beta})
 \left( { m_{\beta} \over m_{\delta}} \right)^{3/2} \right]^{2/3} 
 {m_{\alpha} \over m_{\gamma}} N_{\delta}^{-1/3} \right)\! ,
\label{Oratio}
\end{eqnarray}

\noindent
where eq. (\ref{sim}) has been  taken into account.   
Now, if $m_S \sim m_B$, then we will have

\begin{equation}
{\mid \mbox{\boldmath $F$}^{\beta}_{i_{\alpha}, fluc} \mid \over
\mid \mbox{\boldmath $F$}^{\delta}_{j_{\gamma}, smooth} \mid } = 
 O \left(  \left[ 
 \left( { R_{\delta} \over R_{\beta}} \right)^{2} + (1 - \delta_{\alpha \beta})
 \right]^{2/3}
N_{\delta}^{-1/3} \right)\! ,
\label{Oratio2}
\end{equation}

\noindent
this is indeed a very small quantity. If, on the contrary, background 
and satellite particles have very different masses, eq. (\ref{Oratio})
must be examined more carefully. The most unfavorable case is when
$m_{\delta} \ll m_{\tilde{\delta}}, \gamma = \delta$ and $\alpha = \beta
= \tilde{\delta}$. In this case, eq. (\ref{Oratio}) gives:

\begin{equation}
{\mid \mbox{\boldmath $F$}^{\beta}_{i_{\alpha}, fluc} \mid \over
\mid \mbox{\boldmath $F$}^{\delta}_{j_{\gamma}, smooth} \mid } = 
 O \left(  \left[ 
{ m_{\beta} \over m_{\delta}} \right] ^{4/3}
\left[{ R_{\delta} \over R_{\beta}} \right] ^{4/3} N_{\delta}^{-1/3} \right) ,
\label{Oratio3}
\end{equation}

\noindent
and it would be sufficient to ask that

\begin{equation}
N_{\delta}^{1/3} \gg \left( {m_{\tilde{\delta}} \over m_{\delta}} \right)^{4/3}
\label{gg2}
\end{equation}

\noindent
to ensure that the fluctuating forces are much smaller than the
smooth ones.

Next we compare the timescale for fluctuation, $\tau_{\alpha \beta}$, and
the timescale $T_{\gamma \delta}$ over which the velocity of class 
$\gamma$ particles changes appreciably as a result of the smooth forces 
produced by class $\delta$ particles. The first is of order of the time 
required for the nearest neighbor to travel an interparticle distance,
$\tau_{\alpha \beta} = { 1 \over \sigma n^{1/3}}$, where $\sigma$
is $\sigma_{\alpha}$ or $\sigma_{\tilde{\alpha}}$ and 
$n = n_{\alpha} + (1 - \delta_{\alpha \beta}) n_{\beta}$. As it may happen that
$n_{\alpha}$ and $n_{\beta}$ are very different, we take the maximum
$\tau_{\alpha \beta}$ timescale, corresponding to $n_{min} = N_{min} 
R_{min}^{-3}$. The second timescale is set by $T_{\gamma \delta} =
{ \mid \mbox{\boldmath $v$}_{\gamma} \mid \over \mid \mbox{\boldmath $\nabla$} 
\Phi_{\delta} \mid } \sim {\mid \mbox{\boldmath $v$}_{\gamma} \mid R_{\delta} \over
\sigma^{2}} $, so that we can write

\begin{equation}
{ \tau_{\alpha \beta}^{max} \over T_{\gamma \delta}} \sim
 O \left( N_{min}^{-1/3} {R_{min} \over R_{\delta}} \right) \ll 1,
\label{ll}
\end{equation}

\noindent
where we have used the inequality (\ref{gg}).

The rate of energy exchange between subsystems $B$ and $S$ will be obtained 
as an integral on time which involves the correlation matrix (see below
and BM92 and M93). The correlation matrix is known to fall with time
faster than $t^{-1}$ (Cohen 1975; Kandrup 1980; M93), so that it can be taken
to vanish for times much larger than the fluctuation timescale, and in
particular for times of the order of the macroscopic timescale,
$T_{\gamma \delta}$. As a consequence, in this work we will consider
time intervals after $t = 0$, $\delta t$, which are large as compared with the
fluctuation timescale, $ \tau_{\alpha \beta}^{max}$, but much shorter than
the macroscopic timescale (see Reif 1965 and M93 for a detailed discussion)

\begin{equation}
 \tau_{\alpha \beta}^{max} \ll \delta t \ll min \left[  { \mid \mbox{\boldmath $v$}_{\alpha}
 \mid \over \mid {d \over dt} \mbox{\boldmath $v$}_{\alpha} \mid},
 {\mid \mbox{\boldmath $v$}_{\alpha} \mid \over \mid \mbox{\boldmath $\nabla$} \Phi_{\gamma}
 \mid } \right] .
\label{timeint}
\end{equation}

\subsection{The energy exchange}

The energy of a particle $i_{\alpha}$ is not conserved. Its total instantaneous
variation at time $t$, due to the fluctuating forces caused by class $\beta$
particles is given by:

\begin{equation}
\left( {dE_{i_{\alpha}}^{\beta} \over dt }\right)_{t} =
\mbox{\boldmath $F$}_{i_{\alpha}}^{\beta}(t) \cdot \mbox{\boldmath $v$}_{i_{\alpha}}(t) .
\label{dEi}
\end{equation}

Taking into account the expression for $\mbox{\boldmath $v$}_{i_{\alpha}}(t)$ given
by eq. (\ref{eqmov}), we get for times $ t$ verifying (\ref{timeint})

\begin{equation}
\left( {dE_{i_{\alpha}}^{\beta} \over dt }\right)_{ t} =
\mbox{\boldmath $F$}_{i_{\alpha}}^{\beta}(t) \cdot \left[ \mbox{\boldmath $v$}_{i_{\alpha}, 0 }
+ {1 \over m_{\alpha}} \sum_{\gamma} \int_{0}^{t} dt' 
\mbox{\boldmath $F$}_{i_{\alpha}}^{\gamma} (t') \right],
\label{dEidt}
\end{equation}

\noindent
where the potential terms have been neglected because 
$\mid~\mbox{\boldmath $v$}_{i_{\alpha}, 0 } \mid \gg \mid \mbox{\boldmath $\nabla$} \Phi_{\gamma}
\mid  t $, as ensured by (\ref{timeint}).
The total energy change of the $i_{\alpha}$ particle during the
time interval $(0, \delta t)$ is easily calculated by integrating
eq. (\ref{dEidt}). Summing up on $i_{\alpha}$ we get the
total energy variation of class $\alpha$ particles due to the 
fluctuating forces caused by $\beta$ particles in this time interval:

\begin{eqnarray}
\Delta E_{\alpha, tot}^{\beta } (\delta t) &=&\sum_{i_{\alpha}}
\left. \int_{0}^{\delta t} dt 
\mbox{\boldmath $F$}_{i_{\alpha}}^{\beta}(t) \cdot \right[ 
\mbox{\boldmath $v$}_{i_{\alpha}, 0 } \nonumber \\
&+&\left. {1 \over m_{\alpha}} \sum_{\gamma} \int_{0}^{t} dt'  
\mbox{\boldmath $F$}_{i_{\alpha}}^{\gamma} (t') \right].
\label{deltaEtot}
\end{eqnarray}

We will be interested in the average value of this quantity.

\subsection{Statistical averaging}

The presence and motion of the $i_S$ satellite particle perturbs the 
background. As a result, while the statistical (ensemble) average of the
background fluctuating forces acting on the satellite
 vanishes in the unperturbed state,
$ \langle \mbox{\boldmath $F$}^{B}_{i_S}(t) \rangle _{0} = 0$,
they do not vanish anymore in the real perturbed system,
$ \langle \mbox{\boldmath $F$}^{B}_{i_S}(t) \rangle  \neq 0$.
The same is true  in general for the stochastic forces caused by class $\beta$
particles and acting on class $\alpha$ particles:
$\langle  \mbox{\boldmath $F$}^{\beta}_{i_{\alpha}}(t) \rangle_{0} = 0$, but 
$\langle \mbox{\boldmath $F$}^{\beta}_{i_{\alpha}}(t) \rangle  \neq 0$,
because the phase density and the number of microstates available to 
the system has changed due to the perturbation. Following BM92 and M93,
we assume 
that  the probability of finding a dynamical variable, $Q$, with a 
given value, $P[Q]$, is proportional to the change in the number of
microstates available to the whole system. This change is given by the factor:

\begin{equation}
K = \exp \left [\delta S \right],
\label{defK}
\end{equation}

\noindent
where $\delta S$ is the entropy change of the system as a result of the 
distortion. The expected value of a dynamical function, $Q$, can
now be written as:

\begin{equation}
\left \langle Q \right \rangle= {\int d\Gamma Q(\Gamma) f_{0}(\Gamma) K
\over \int d\Gamma  f_{0}(\Gamma) K} ,
\label{aver}
\end{equation}

\noindent
where $\Gamma$ are the variables defining the phase space of the $N_B + N_S$
particles and $ f_{0}(\Gamma)$ is the distribution function for the
unperturbed state. 

As a result of the fluctuations, a generic particle initially with energy
$\varepsilon_{i_{\alpha}}$, will change to an energy 
$\varepsilon_{i_{\alpha}}' = \varepsilon_{i_{\alpha}}+
\delta \varepsilon_{i_{\alpha}}$, $\alpha = B,S$.
Recalling the definition of  entropy:

\begin{equation}
S = S_0 - \int f \ln f d\Gamma,
\label{defS}
\end{equation}

\noindent
with $S_0$  a constant, this energy change results into a total entropy
variation given by:

\begin{equation}
\delta S =\sum_{\alpha} \left[ - \sum_{i_{\alpha}} \ln 
f^{\alpha}(\varepsilon_{i_{\alpha}}') + \sum_{i_{\alpha}} \ln
f^{\alpha}(\varepsilon_{i_{\alpha}})\right],
\label{varS}
\end{equation}

\noindent
where $f^{\alpha}(\varepsilon_{i_{\alpha}})$ is the class $\alpha$ one
particle distribution function for the unperturbed state corresponding
to an energy $\varepsilon_{i_{\alpha}}$. Most two-body encounters are weak,
and then $\delta \varepsilon_{i_{\alpha}} \ll \varepsilon_{i_{\alpha}}$.
Expanding the r.h.s. of eq.(\ref{varS}) and then the exponential in 
eq.(\ref{defK}), we obtain that the change in the probability function is
given approximately by:

\begin{equation}
K = 1 - \sum_{\alpha} \sum_{i_{\alpha}} {1 \over f^{\alpha}}
{\partial f^{\alpha} \over \partial \varepsilon_{i_{\alpha}}} 
\delta \varepsilon_{i_{\alpha}}.
\label{expK}
\end{equation}

For most astronomical applications it is justified to use an isothermal 
Maxwellian one-particle distribution function,

\begin{equation}
f_{0}^{\alpha}(\mbox{\boldmath $r$},\mbox{\boldmath $u$}) = 
{n_{\alpha}(\mbox{\boldmath $r$}) \over N_{\alpha}
\left( 2 \pi \sigma_{\alpha}^2 \right)^{3/2}}  
\exp \left [ -\beta_{\alpha}
 \varepsilon_{i_{\alpha}} \right ],
\label{fMax}
\end{equation}

\noindent
where $\mbox{\boldmath $u$}$ is the velocity vector,
$\sigma_{\alpha}$ is the velocity dispersion of class $\alpha$
particles and $\beta_{\alpha} = {1 \over m_{\alpha} \sigma_{\alpha}^2}$
is an inverse temperature.
Introducing this expression for the distribution function in eq. (\ref{expK}),
the factor of change becomes:

\begin{equation}
K = 1 + \sum_{\alpha} \beta_{\alpha} \sum_{i_{\alpha}} 
\delta \varepsilon_{i_{\alpha}} .
\label{Kfinal}
\end{equation}

The next step is to find out an expression for the energy variation. 
The total energy of the system must be conserved, so that we have:

\begin{equation}
\sum_{\alpha} \sum_{i_{\alpha}} \delta \varepsilon_{i_{\alpha}} = 0.
\label{Econs}
\end{equation}

Moreover, for one given particle, $i_{\alpha}$, its energy variation can
be expressed as:

\begin{equation}
\delta \varepsilon_{i_{\alpha}} = \Delta E_{i_{\alpha}}^{\alpha} + \Delta
E_{i_{\alpha}}^{\tilde{\alpha}} - a_{i_{\alpha}} \Delta E_{\alpha, tot}^{\alpha}
- b_{i_{\alpha}} \Delta E_{\tilde{\alpha}, tot}^{\alpha},
\label{deltE}
\end{equation}

\noindent
where  
  $a_{i_{\alpha}}$
 is the fraction of the total class $\alpha$ autointeraction energy,
 $\Delta E_{\alpha, tot}^{\alpha}$, absorbed by particle $i_{\alpha}$,
and $b_{i_{\alpha}}$ is the fraction of the total interaction energy 
between class $\alpha$ and $\tilde{\alpha}$ particles, caused by the 
fluctuating forces of the $\alpha$ subsystem, absorbed by  $i_{\alpha}$
particle. 
Summing on $i_{\alpha}$ in eq. (\ref{deltE}), 
we get the total energy change of subsystem $\alpha$:

\begin{equation}
\sum_{i_{\alpha}} \delta \varepsilon_{i_{\alpha}} =
 \Delta E_{\alpha, tot}^{\tilde{\alpha}} - 
 \Delta E_{\tilde{\alpha}, tot}^{\alpha},
\label{sumE}
\end{equation}

\noindent
compatible with energy conservation (eq. (\ref{Econs})). To write down eq. 
(\ref{sumE}) from (\ref{deltE}), 
it has been taken into account that $\sum_{i_{\alpha}}
a_{i_{\alpha}} = \sum_{i_{\alpha}} b_{i_{\alpha}} = 1$. Energy signs are
such that $\Delta E_{i_{\alpha}}^{\gamma} > 0$ if particle $i_{\alpha}$ 
gains energy due to the fluctuating forces caused by class $\gamma$
particles and conversely. With this convention, if 
 energy flux is from subsystem $\alpha$
to subsystem $\tilde{\alpha}$, 
then   $ \Delta E_{\alpha, tot}^{\tilde{\alpha}} < 0$
(subsystem $\alpha$ loses energy), and $ \Delta 
E_{\tilde{\alpha}, tot}^{\alpha}>0$
(subsystem $\tilde{\alpha}$ gains energy), so that 
$\sum_{i_{\alpha}} \delta \varepsilon_{i_{\alpha}} < 0$ and
$\sum_{i_{\tilde{\alpha}}} \delta \varepsilon_{i_{\tilde{\alpha}}} > 0$.
Inserting eq. (\ref{sumE}) in eq. (\ref{Kfinal}) and recalling eq. 
(\ref{deltaEtot}), we obtain:

\begin{eqnarray}
 K\!\! &\! =\! &\!\! 1 + \sum_{\alpha} \beta_{\alpha}
\left\{ \sum_{i_{\alpha}} \int_{0}^{\delta t} dt
\mbox{\boldmath $F$}_{i_{\alpha}}^{\tilde{\alpha}} (t) \cdot 
\right[ \mbox{\boldmath $v$}_{i_{\alpha},0} \nonumber \\
\!\!&\!+\!&\!\! {1 \over m_{\alpha}} \sum_{\gamma} \int_{0}^{ t} dt' 
\mbox{\boldmath $F$}_{i_{\alpha}}^{\gamma} (t') 
- \left. \sum_{i_{\tilde{\alpha}}} \int_{0}^{\delta t} dt
\mbox{\boldmath $F$}^{\alpha}_{i_{\tilde{\alpha}}} (t) \cdot \right[ 
\mbox{\boldmath $v$}_{i_{\tilde{\alpha}},0} \nonumber \\
\!\!&\!+\!&\!\! \left. \left. {1 \over m_{\tilde{\alpha}}} 
\sum_{\gamma} \int_{0}^{t} dt' 
\mbox{\boldmath $F$}_{i_{\tilde{\alpha}}}^{\gamma} (t') \right] \right\} . 
\label{Klargo}
\end{eqnarray}

Eq. (\ref{aver}) allows us now to
write the ensemble average of the instantaneous
energy variation at time $t$ of a generic class $\mu$ particle due
to the stochastic forces caused by class $\nu$ particles:

\begin{eqnarray}
&&\!\!\!\!\!\!\!\!\!\!\!\!\left\langle \left(  {dE_{i_{\mu}}^{\nu} \over dt} \right)_{\delta t} 
\right\rangle= 
\left. \left. \int d\Gamma f_{0}({\Gamma})\right[ 
\mbox{\boldmath $F$}_{i_{\mu}}^{\nu}(\delta t)
 \cdot \right( \mbox{\boldmath $v$}_{i_{\mu},0} \nonumber \\
&&\!\!\!\!\!\!\!+ \left. \left. \!{1 \over m_{\mu}} \sum_{\rho} 
\int_{0}^{\delta  t} \!\!\!dt 
\mbox{\boldmath $F$}_{i_{\mu}}^{\rho} (t) \right) \right]
\left[  1 + \sum_{\alpha} \beta_{\alpha}
\left\{ \sum_{i_{\alpha}} \int_{0}^{\delta t} \!\!dt
\mbox{\boldmath $F$}_{i_{\alpha}}^{\tilde{\alpha}} (t)  \right. \right. 
\nonumber\\
&&\!\!\!\!\!\!\!\cdot\left( \mbox{\boldmath $v$}_{i_{\alpha},0} +
{1 \over m_{\alpha}} \sum_{\gamma} \int_{0}^{ t}\!\! dt' 
\mbox{\boldmath $F$}_{i_{\alpha}}^{\gamma} (t') \right) -
\sum_{i_{\tilde{\alpha}}} \int_{0}^{\delta t} \!\!dt
\mbox{\boldmath $F$}^{\alpha}_{i_{\tilde{\alpha}}} (t)  \nonumber \\
&&\!\!\!\!\!\!\! 
\cdot\left( \mbox{\boldmath $v$}_{i_{\tilde{\alpha}},0} + 
\left.\left. {1 \over m_{\tilde{\alpha}}}
\sum_{\gamma} \int_{0}^{ t} \!\!dt' 
\mbox{\boldmath $F$}_{i_{\tilde{\alpha}}}^{\gamma} (t') \right) 
\right\}\right] \left[ \int d\Gamma f_{0}(\Gamma) K \right]^{-1}
\label{dEaver}
\end{eqnarray}

To second order in the fluctuating forces,  taking into account that the
average of the stochastic forces in the unperturbed state vanishes, 
and summing on the $i_{\mu}$
subindex to obtain the global effect,
we get:

\begin{eqnarray}
\left\langle \left( {dE_{\mu, tot}^{\nu} \over dt} \right)_{\delta t} 
\right\rangle  =
\sum_{i_{\mu}}{1 \over m_{\mu}} \left[
\int_{-\delta t}^{0} ds Tr \left[ C_{i_{\mu} i_{\mu}}^{\nu \nu} (s) 
\right] \right. \nonumber \\
+
\left. (\beta_{\tilde{\nu}} - \beta_{\nu}) 
\sum_{j_{\tilde{\nu}}} \int_{-\delta t}^{0} \!\!ds
\mbox{\boldmath $v$}_{i_{\mu}, 0} C_{i_{\mu} j_{\tilde{\nu}}}^{\nu \nu} (s) 
\mbox{\boldmath $v$}_{j_{\tilde{\nu}}, 0} \right] ,
\label{dEgen}
\end{eqnarray}

\noindent
where

\begin{equation}
C_{i_{\gamma} j_{\delta}}^{\alpha \beta} (s) = 
\langle \mbox{\boldmath $F$}_{i_{\gamma}}^{\alpha} (0)
\otimes \mbox{\boldmath $F$}_{ j_{\delta}}^{\beta} (s) \rangle_{0}
\label{defC}
\end{equation}

\noindent
is the correlation matrix (see Appendix A; the symbol $\otimes$
stands for the tensorial product of vectors 
$\mbox{\boldmath $F$}_{i_{\gamma}}^{\alpha}$
and $\mbox{\boldmath $F$}_{ j_{\delta}}^{\beta}$, the average is on the
unperturbed states of particles $\alpha$ and $\beta$ and the invariance of the 
correlation matrix under time translation has been taken into account). 
Because in the unperturbed state
two different particles are not correlated, the correlation matrix defined
in the previous equation vanishes if $\alpha \neq \beta$. This has been taken 
into account to deduce eq. (\ref{dEgen}) from eq. (\ref{dEaver}). Note that
because for $s \ge \delta t$ the correlations vanish, the lower limit
of the integrals can be extended up to $-\infty$.
Eq. (\ref{dEgen}) is the expression of
the global instantaneous energy variation of subsystem $\mu$ due to the
fluctuating forces of subsystem $\nu$. 
The integrand in the first term of the r.h.s. of eq. 
(\ref{dEgen}) is invariant under time reversal. Extending the 
integral to positive $s$, this term is the  corresponding
power spectrum at zero frequency 
(Wiener-Khintchine theorem, see Reif 1965 and BM92)
and, consequently,  it  is positive 
and represents a heating term of class $\mu$ particles due to the fluctuating
forces caused by $\nu$ particles.
It is of order $O({1 \over N}) \ll 1$ relative to the second term.
Terms of this kind will be neglected in this work.
The second term can be either positive or negative, depending on the sign
of $(\beta_{\tilde{\nu}} - \beta_{\nu})$. In the  next section it will be 
explicitly calculated for $\mu, \nu = B, S$.

\section{The instantaneous energy variation}

\subsection{Energy exchange between the satellite and the background}

Energy flows between the satellite and the background as a  result
of either the global energy variation of the satellite particles due to
the fluctuating forces of the background, or, conversely, 
the global energy variation of the
background particles due to the fluctuating forces of the satellite.
These fluxes are given by eq. (\ref{dEgen}) when   $\mu \neq \nu$.

Let us first calculate the effect due to the stochastic forces of the
background. This term could cause an energy flux responsible for the
satellite deacceleration. Eq. (\ref{dEgen}) with $\mu = S, \nu = B$ and
neglecting the heating term, gives:

\begin{eqnarray}
&&\!\!\!\!\!\!\!\!\!\!\!\! \left\langle \left( {dE_{S, tot}^{B} \over dt} 
\right)_{\delta t} \right\rangle =\nonumber \\
&& = (\beta_{S} - \beta_{B}) \sum_{i_{S}} 
\sum_{j_{S}} \int_{-\infty}^{0} ds \mbox{\boldmath $v$}_{i_{S},0} 
C_{i_{S} j_{S}}^{B B} (s) \mbox{\boldmath $v$}_{j_{S},0},
\label{EBS}
\end{eqnarray} 

\noindent
where the correlation matrix is given by eq. (\ref{A10}) with $\alpha = B$
and $\gamma = \delta = S$. In the case of a point-like satellite with
$m_S \gg m_B$, $\beta_{S} \ll \beta_{B}$ and then the satellite loses energy to
the background. In our case, however, it is not excluded in principle
that  $\beta_{S} > \beta_{B}$ and then the energy would 
flow from the background
to the satellite.

The integral over $ds$ in eq. (\ref{EBS}) can be calculated
taking into  account eq. (\ref{A10}) and
the equality:

\begin{eqnarray}
\int_{-\infty}^{0} ds {  \mbox{\boldmath $A$} +  \mbox{\boldmath $V$}  s
\over \mid  \mbox{\boldmath $A$} +  \mbox{\boldmath $V$}  s \mid^{3}} =
{ \mbox{\boldmath $A$} \cdot \mbox{\boldmath $V$}  + A V \over
V^{2} A_{\perp}^{2} } 
 \left[ { \mbox{\boldmath $A$} \over A} -
  { \mbox{\boldmath $V$} \over V} \right]
\label{intF}
\end{eqnarray}

\noindent
where $ \mbox{\boldmath $A$}_{\perp}$ is the projection of vector $ \mbox{\boldmath $A$}$ on the plane
normal to vector $\mbox{\boldmath $V$} $. In our case either
$\mbox{\boldmath $V$} \equiv \mbox{\boldmath $v$}_{j_{S}} - 
\mbox{\boldmath $v$}_{h_{B}}$ or $\mbox{\boldmath $V$} \equiv 
\mbox{\boldmath $v$}_{j_{S}}$, (see eq. (\ref{A10})), and 
$\mbox{\boldmath $A$} \equiv \mbox{\boldmath $r$}_{j_{S}} - 
\mbox{\boldmath $r$}_{h_{B}}$ or
$\mbox{\boldmath $A$} \equiv \mbox{\boldmath $r$}_{j_{S}} - 
\mbox{\boldmath $r$}^{ '}$. We obtain:

\begin{eqnarray}
&&\!\!\!\!\!\!\!\!\!\!\!\!
\left\langle \left( {dE_{S, tot}^{B} \over dt} 
\right)_{\delta t} \right\rangle  =
G^{2} m_{B}^{2} m_{S}^{2} (\beta_{S} - \beta_{B}){N_{B}^{'} \over N_{B}} 
\nonumber \\
&&
\times \sum_{i_{S}} \sum_{j_{S}} \left[ \sum_{h_{B}}  \left\{ 
{ (\mbox{\boldmath $r$}_{i_{S}} - \mbox{\boldmath $r$}_{h_{B}}) \cdot \mbox{\boldmath $v$}_{i_{S}} \over
\mid \mbox{\boldmath $r$}_{i_{S}} - \mbox{\boldmath $r$}_{h_{B}} \mid^{3}} \right\}
\right. \nonumber \\
&& 
\times \left\{ { [ \mbox{\boldmath $v$}_{j_{S}} 
\cdot (\mbox{\boldmath $r$}_{j_{S}} - \mbox{\boldmath $r$}_{h_{B}})]
[(\mbox{\boldmath $r$}_{j_{S}} - \mbox{\boldmath $r$}_{h_{B}}) 
\cdot (\mbox{\boldmath $v$}_{j_{S}} -
\mbox{\boldmath $v$}_{h_{B}})] \over
\mid \mbox{\boldmath $r$}_{j_{S}} - \mbox{\boldmath $r$}_{h_{B}} \mid \mid \mbox{\boldmath $v$}_{j_{S}} -
\mbox{\boldmath $v$}_{h_{B}} \mid^{2} 
\mid ( \mbox{\boldmath $r$}_{j_{S}} - 
\mbox{\boldmath $r$}_{h_{B}})_{\perp} \mid^{2}}  \right.\nonumber\\
&&
-{\mid \mbox{\boldmath $r$}_{j_{S}} - 
\mbox{\boldmath $r$}_{h_{B}} \mid^{2}
(\mbox{\boldmath $v$}_{j_{S}} -\mbox{\boldmath $v$}_{h_{B}}) \cdot 
\mbox{\boldmath $v$}_{j_{S}} \over
\mid \mbox{\boldmath $r$}_{j_{S}} - \mbox{\boldmath $r$}_{h_{B}} \mid \mid \mbox{\boldmath $v$}_{j_{S}} -
\mbox{\boldmath $v$}_{h_{B}} \mid^{2} 
\mid ( \mbox{\boldmath $r$}_{j_{S}} - 
\mbox{\boldmath $r$}_{h_{B}})_{\perp} \mid^{2}} \nonumber\\
&&
+ \left. { \mbox{\boldmath $v$}_{j_{S}} \cdot 
( \mbox{\boldmath $r$}_{j_{S}} - \mbox{\boldmath $r$}_{h_{B}})_{\perp} \over
\mid \mbox{\boldmath $v$}_{j_{S}} - \mbox{\boldmath $v$}_{h_{B}} \mid 
\mid ( \mbox{\boldmath $r$}_{j_{S}} - \mbox{\boldmath $r$}_{h_{B}})_{\perp} \mid^{2}} \right\} \nonumber \\
&& 
+ \left. \!\!{1 \over N_{B}} \!\!\int\!\! 
d\mbox{\boldmath $r$} d\mbox{\boldmath $r$}' 
n_{B} (\mbox{\boldmath $r$})
 n_{B} (\mbox{\boldmath $r$}') { [ (\mbox{\boldmath $r$}_{i_{S}} - 
\mbox{\boldmath $r$}) \cdot \mbox{\boldmath $v$}_{i_{S}}]
\over \mid \mbox{\boldmath $r$}_{i_{S}} - \mbox{\boldmath $r$} 
\mid^{3} \mid \mbox{\boldmath $r$}_{j_{S}} 
 - \mbox{\boldmath $r$}' \mid } \right] 
\label{monsterE}
\end{eqnarray}

\noindent
where the sum on a generic subindex, $i_{\alpha}$, means $N_{\alpha}$ times
the average on velocity and positions of class $\alpha$ particles, and is
carried out by means of the distribution function given in eq. (\ref{fMax}).

To proceed further, we recall that the velocity distribution of
background particles is assumed to be 
Maxwellian with zero mean and dispersion $\sigma_B$,
and the velocity distribution of the satellite particles has mean equal to
the satellite center of mass velocity, $\mbox{\boldmath $v$}_{CMS}$, and dispersion
$\sigma_S$. The integration over $d \mbox{\boldmath $v$}_{h_{B}}$ gives:

\begin{eqnarray}
&&\!\!\!\!\!\!\!\!\!\!
\left\langle  \left(  {dE_{S, tot}^{B} \over dt} \right)_{\delta t} 
\right\rangle =
G^{2} m_{B}^{2} m_{S}^{2} (\beta_{S} - \beta_{B}) \nonumber \\
&&\times \sum_{i_{S}} \sum_{j_{S}} \left[ \int d\mbox{\boldmath $r$}_{h_{B}} 
n_{B}(\mbox{\boldmath $r$}_{h_{B}})
\left\{
{ (\mbox{\boldmath $r$}_{i_{S}} - \mbox{\boldmath $r$}_{h_{B}}) \cdot 
\mbox{\boldmath $v$}_{i_{S}} \over
\mid \mbox{\boldmath $r$}_{i_{S}} - \mbox{\boldmath $r$}_{h_{B}} 
\mid^{3}} \right\}
\right. \nonumber \\
& & 
\times \left\{ {\exp(\gamma_{j_{S}}^{2} - y_{j_{S}}^{2}) 
\mbox{erf}(\gamma_{j_{S}}) - 1
\over \mid \mbox{\boldmath $r$}_{j_{S}} - \mbox{\boldmath $r$}_{h_{B}} 
\mid} \right\} \nonumber \\
& & 
\left. +\! {1 \over N_{B}} \!\!\int\!\! d\mbox{\boldmath $r$} 
d\mbox{\boldmath $r$}' n_{B} (\mbox{\boldmath $r$})
n_{B} (\mbox{\boldmath $r$}') { [ (\mbox{\boldmath $r$}_{i_{S}} - 
\mbox{\boldmath $r$}) \cdot \mbox{\boldmath $v$}_{i_{S}}]
\over \mid \mbox{\boldmath $r$}_{i_{S}} - \mbox{\boldmath $r$} \mid^{3} 
\mid \mbox{\boldmath $r$}_{j_{S}}
- \mbox{\boldmath $r$}' \mid } \right] 
\label{EBScal}
\end{eqnarray}

\noindent
where $\mbox{erf}$ stands for the error function, 
$\mbox{\boldmath $y$}_{j_{S}} \equiv 
{\mbox{\boldmath $v$}_{j_{S}} \over \sqrt{2} \sigma_{B}}$,
$\gamma_{j_{S}} \equiv \mbox{\boldmath $y$}_{j_{S}} \cdot 
\mbox{\boldmath $\zeta$} $ with $\mbox{\boldmath $\zeta$} \equiv
{(\mbox{\boldmath $r$}_{j_{S}} - \mbox{\boldmath $r$}_{h_{B}}) 
\over \mid \mbox{\boldmath $r$}_{j_{S}} - 
\mbox{\boldmath $r$}_{h_{B}}\mid}$ and taking $\frac{N_{B}^{'}}{N_B}=1$.

The integration over $d \mbox{\boldmath $v$}_{j_{S}}$ involves the integral:

\begin{equation}
I = \int d \mbox{\boldmath $v$}_{j_{S}} f_{M}(\mbox{\boldmath $v$}_{j_{S}}) 
\exp(\gamma_{j_{S}}^{2} - y_{j_{S}}^{2}) \mbox{erf}(\gamma_{j_{S}}).
\label{defI}
\end{equation}

In Appendix B we show that

\begin{equation}
 I = { \sigma_{B}^{2} \over \sigma_{T}^{2}} 
 \exp (\alpha_{CMS}^{2} - x_{CMS}^{2}) \mbox{erf} (\alpha_{CMS}),
\label{resI}
\end{equation}

\noindent
where $\sigma_{T}^{2} = \sigma_{B}^{2} +  \sigma_{S}^{2}$,
 $\mbox{\boldmath $x$}_{CMS} \equiv {\mbox{\boldmath $v$}_{CMS} \over 
\sqrt{2} \sigma_{T}}$
and $\alpha_{CMS} \equiv \mbox{\boldmath $x$}_{CMS} \cdot 
\mbox{\boldmath $\zeta$}$.
 The integration
over $d \mbox{\boldmath $v$}_{i_{S}}$ is trivial recalling that 
$\mbox{\boldmath $v$}_{i_{S}} = \mbox{\boldmath $v$}_{CMS} + \mbox{\boldmath 
$u$}_{i_{S}}$ and that for a
Maxwellian distribution the contribution of $\mbox{\boldmath $u$}_{i_{S}}$ to
the first moments vanishes. Taking $m_{\alpha} n_{\alpha} = 
\rho_{\alpha}$, we finally get:

\begin{eqnarray}
&&\!\!\!\!\!\!\!\!\!\!\!\!\!\left\langle\!\! \left(\!{dE_{S, tot}^{B} \over dt} 
\right)_{\!\delta t} 
\right\rangle \!\!=\!\!
 \left(\!  {m_B \sigma_{B}^{2} \over m_S \sigma_{S}^{2}} \!-\! 1 \!\right)
{G^{2} \over \sigma_{B}^{2}}\!\! 
\int\!\! d\mbox{\boldmath $r$}_{i_{S}} d\mbox{\boldmath $r$}_{j_{S}}
\rho_{S} (\mbox{\boldmath $r$}_{i_{S}}) \rho_{S} 
(\mbox{\boldmath $r$}_{j_{S}}) \nonumber \\
&&\!\!\times\left[
  \int d\mbox{\boldmath $r$}_{h_{B}} \rho_{B}(\mbox{\boldmath $r$}_{h_{B}}) 
\left\{
{ (\mbox{\boldmath $r$}_{i_{S}} - \mbox{\boldmath $r$}_{h_{B}}) \cdot \mbox{\boldmath $v$}_{CMS} \over
\mid \mbox{\boldmath $r$}_{i_{S}} - \mbox{\boldmath $r$}_{h_{B}} \mid^{4}} 
\right\} \right.\nonumber \\
&&\!\! \times \left\{  {\sigma_{B}^{2} \over \sigma_{T}^{2}}
exp(\alpha_{CMS}^{2} - x_{CMS}^{2}) \mbox{erf}(\alpha_{CMS}) - 1
\right\} \nonumber \\
&&\!\! +\left. 
{1 \over N_{B}} \int d\mbox{\boldmath $r$} d\mbox{\boldmath $r$}' 
\rho_{B} (\mbox{\boldmath $r$})
n_{B} (\mbox{\boldmath $r$}') 
{ [ (\mbox{\boldmath $r$}_{i_{S}} - \mbox{\boldmath $r$}) \cdot \mbox{\boldmath $v$}_{CMS}]
\over \mid \mbox{\boldmath $r$}_{i_{S}} - \mbox{\boldmath $r$} \mid^{3} \mid \mbox{\boldmath $r$}_{j_{S}}
- \mbox{\boldmath $r$}' \mid } \right] 
\label{EBSfin}
\end{eqnarray}

A specification of the density distribution of both the satellite and
the background  is needed in order to carry out the integrals over the space 
variables.

Next we calculate the variation of the background energy caused by the 
stochastic forces of the satellite. It can be obtained from eq.(\ref{dEgen})
with $\mu = B$ and $\nu = S$, neglecting the heating term: 

\begin{eqnarray}
&&\!\!\!\!\!\!\!\!\!\!\left\langle  \left(  {dE_{B, tot}^{S} \over dt} \right)_{\delta t} \right\rangle  =\nonumber \\
&&=(\beta_{B} - \beta_{S}) \sum_{i_{B}} \sum_{j_{B}} \int_{-\infty}^{0}
ds \mbox{\boldmath $v$}_{i_{B},0} C_{i_{B} j_{B}}^{S S} (s) \mbox{\boldmath $v$}_{j_{B},0}.
\label{ESB}
\end{eqnarray}

For a background at rest with a Maxwellian velocity distribution function,
this energy rate vanishes when one performs the integration over 
$d \mbox{\boldmath $v$}_{i_{B}}$. We conclude that the effect of fluctuating forces of
the satellite acting on the background only heat it, and have no effect on
a variation of the orbital energy of either the satellite or the background.

The total instantaneous energy flow between the satellite and the background 
is given by the difference between the l.h.s. of the eqs. (\ref{EBSfin}) and
(\ref{ESB}) (see eq. (\ref{sumE})). It can
be written in terms of the rate of change of the satellite
orbital and internal energies as:

\begin{eqnarray}
&&\!\!\!\!\!\!\!\!\!\!\left\langle  \left(  {dE_{ tot}^S \over dt} \right)_{\delta t} \right\rangle=
\left\langle  \left(  {dE_{orb}^{S} \over dt} \right)_{\delta t} \right\rangle +
\left\langle  \left(  {dE_{in}^{S} \over dt} \right)_{\delta t} \right\rangle 
\nonumber \\
&&=\left\langle  \left(  {dE_{S, tot}^{B} \over dt} \right)_{\delta t} \right\rangle 
\label{flow}
\end{eqnarray}

\noindent
where $E_{orb}^{S}$ and $E_{in}^{S}$ stand for the orbital and internal
energy of the satellite, respectively, and the second equality results
from the zero value of the flow given by eq. (\ref{ESB}).
The rate of change of the satellite internal energy can be calculated with
the help of eq. (\ref{aver}) with $ Q = {dE_{in}^{S} \over dt}$ or:

\begin{equation}
Q = m_{S} \sum_{i_{S}} \mbox{\boldmath $u$}_{i_{S}} {d \over dt} \mbox{\boldmath $u$}_{i_{S}} ,
\label{opEin}
\end{equation}

\noindent
where $\mbox{\boldmath $u$}_{i_{S}} = \mbox{\boldmath $v$}_{i_{S}} - \mbox{\boldmath $v$}_{CMS}$ is the velocity
of $i_{S}$ particles with respect to its center of mass. This gives an
expression similar to eq. (\ref{EBSfin}), except that now $ \mbox{\boldmath $v$}_{CMS}$
takes a zero value, so that

\begin{equation}
\left\langle  \left(  {dE_{in}^{S} \over dt} \right)_{\delta t} \right\rangle = 0 ,
\label{Ein}
\end{equation}

\noindent
and then eq. (\ref{EBSfin}) gives, at second order in the
fluctuations, the rate of change of the satellite {\it orbital} energy.

\subsection{The self-interaction energies of the background and the 
satellite}

When $\mu = \nu = B$ or $S$, eq. (\ref{dEgen}) describes the instantaneous
energy variation rate of class $\mu$ particles due to the stochastic forces 
caused by $\mu$ particles themselves. 

According to the eq. (\ref{sumE}) these energies do not play any role in the
energy change of the subsystems S or B. They only represent the energy 
change of an individual particle (see eq. (\ref{deltE})).

The self-interaction energy of the background is easily obtained from 
 eq. (\ref{dEgen}) with $\mu = \nu = B$:

\begin{eqnarray}
&&\!\!\!\!\!\!\!\!\!\!\left\langle  \left(  {dE_{B, tot}^{B} \over dt} \right)_{\delta t} \right\rangle  =\nonumber \\
&&=(\beta_{S} - \beta_{B}) \sum_{i_{B}} \sum_{j_{S}} \int_{-\infty}^{0}
ds \mbox{\boldmath $v$}_{i_{B},0} C_{i_{B} j_{S}}^{B B} (s) \mbox{\boldmath $v$}_{j_{S},0}.
\label{EBB}
\end{eqnarray}

Again, the integration over $d \mbox{\boldmath $v$}_{i_{B}}$ makes it vanish for a
background at rest with an isotropic velocity distribution function:
because no translation energy is available, the autointeraction results only
in a slow heating of the background.

Regarding the autointeraction energy of the satellite,  eq. (\ref{dEgen})
gives:

\begin{eqnarray}
&&\!\!\!\!\!\!\!\!\!\!\left\langle  \left(  {dE_{S, tot}^{S} \over dt} \right)_{\delta t} \right\rangle  =\nonumber \\
&&=(\beta_{B} - \beta_{S}) \sum_{i_{S}} \sum_{j_{B}} \int_{-\infty}^{0}
ds \mbox{\boldmath $v$}_{i_{S},0} C_{i_{S} j_{B}}^{S S} (s) \mbox{\boldmath $v$}_{j_{B},0}.
\label{ESS}
\end{eqnarray}

The correlation matrix is given by eq. (\ref{A10}) with $\alpha = S, \gamma =
S$ and $\delta = B$. Substituting 
the expression for $ C_{i_{S} j_{B}}^{S S} (s)$
 in eq. (\ref{ESS}), and performing the integral over $ds$ with the help
 of eq. (\ref{intF}), we obtain:

\begin{eqnarray}
&&\!\!\!\!\!\!\!\!\!\!\left\langle 
\left(  {dE_{S, tot}^{S} \over dt} \right)_{\delta t} \right\rangle  =
G^{2} m_{B}^{2} m_{S}^{2} (\beta_{B} - \beta_{S}){N_{S}^{'} \over N_{S}}
\nonumber \\
&&
\times \sum_{i_{S}} \sum_{j_{B}} \left[ \sum_{h_{S}}  \left\{
{ (\mbox{\boldmath $r$}_{i_{S}} - \mbox{\boldmath $r$}_{h_{S}}) \cdot \mbox{\boldmath $v$}_{i_{S}} \over
\mid \mbox{\boldmath $r$}_{i_{S}} - \mbox{\boldmath $r$}_{h_{S}} \mid^{3}} 
\right\} \right. \nonumber \\
&& \times \left\{ { [ \mbox{\boldmath $v$}_{j_{B}} \cdot 
(\mbox{\boldmath $r$}_{j_{B}} - \mbox{\boldmath $r$}_{h_{S}})]
[(\mbox{\boldmath $r$}_{j_{B}} - \mbox{\boldmath $r$}_{h_{S}}) 
\cdot (\mbox{\boldmath $v$}_{j_{B}} - \mbox{\boldmath $v$}_{h_{S}})] 
\over
\mid \mbox{\boldmath $r$}_{j_{B}} - \mbox{\boldmath $r$}_{h_{S}} \mid 
\mid \mbox{\boldmath $v$}_{j_{B}} -
\mbox{\boldmath $v$}_{h_{S}} \mid^{2}
\mid ( \mbox{\boldmath $r$}_{j_{B}} -
\mbox{\boldmath $r$}_{h_{S}})_{\perp} \mid^{2}} 
\right. \nonumber \\
&& - 
{\mid \mbox{\boldmath $r$}_{j_{B}} - \mbox{\boldmath $r$}_{h_{S}} \mid^{2}
(\mbox{\boldmath $v$}_{j_{B}} -\mbox{\boldmath $v$}_{h_{S}}) \cdot 
\mbox{\boldmath $v$}_{j_{B}} 
\over
\mid \mbox{\boldmath $r$}_{j_{B}} - \mbox{\boldmath $r$}_{h_{S}} \mid 
\mid \mbox{\boldmath $v$}_{j_{B}} -
\mbox{\boldmath $v$}_{h_{S}} \mid^{2}
\mid ( \mbox{\boldmath $r$}_{j_{B}} -
\mbox{\boldmath $r$}_{h_{S}})_{\perp} \mid^{2}} 
\nonumber\\
&& + \left. { \mbox{\boldmath $v$}_{j_{B}} \cdot
 ( \mbox{\boldmath $r$}_{j_{B}} - \mbox{\boldmath $r$}_{h_{S}})_{\perp} \over
 \mid \mbox{\boldmath $v$}_{j_{B}} - \mbox{\boldmath $v$}_{h_{S}} \mid
 \mid ( \mbox{\boldmath $r$}_{j_{B}} - \mbox{\boldmath $r$}_{h_{S}})_{\perp} \mid^{2}} \right\} \nonumber \\
&& \left. + {1 \over N_{S}} \!\!\int \!d\mbox{\boldmath $r$} d\mbox{\boldmath $r$}' n_{S} (\mbox{\boldmath $r$})
  n_{S} (\mbox{\boldmath $r$}') { [ (\mbox{\boldmath $r$}_{i_{S}} - \mbox{\boldmath $r$}) \cdot \mbox{\boldmath $v$}_{i_{S}}]
   \over \mid \mbox{\boldmath $r$}_{i_{S}} - \mbox{\boldmath $r$} \mid^{3} \mid \mbox{\boldmath $r$}_{j_{B}}
    - \mbox{\boldmath $r$}' \mid } \right] .
\label{monESS}
\end{eqnarray}

The integrations over velocities can be carried out following the same steps
as in the previous section and we finally have:

\begin{eqnarray}
&&\!\!\!\!\!\!\!\!\!\!\!\!\!\!
\left\langle\!\! \left( \!{dE_{S, tot}^{S} \over dt} 
\right)_{\!\delta t} \right\rangle  \!=\!
\left( \! {m_S \sigma_{S}^{2}
\over m_B \sigma_{B}^{2}} \!-\! 1 \!\right)
{G^{2} \over \sigma_{S}^{2}} \!\! \int \!\!
d\mbox{\boldmath $r$}_{i_{S}} d\mbox{\boldmath $r$}_{j_{B}}
\rho_{S} (\mbox{\boldmath $r$}_{i_{S}}) \rho_{B} 
(\mbox{\boldmath $r$}_{j_{B}})\!\! \nonumber \\
&& 
\!\!\times \left[- {\sigma_{B}^{2} \over \sigma_{T}^{2}}
  \int d\mbox{\boldmath $r$}_{h_{S}} \rho_{S}(\mbox{\boldmath $r$}_{h_{S}})
  \left\{
  { (\mbox{\boldmath $r$}_{i_{S}} - \mbox{\boldmath $r$}_{h_{S}}) \cdot \mbox{\boldmath $v$}_{CMS} \over
  \mid \mbox{\boldmath $r$}_{i_{S}} - \mbox{\boldmath $r$}_{h_{S}} \mid^{3}} 
\right\}  \right. \nonumber \\
&& \!\!\times \left\{ {exp(\alpha_{CMS}^{2} - x_{CMS}^{2}) \mbox{erf}(\alpha_{CMS})
  \over \mid \mbox{\boldmath $r$}_{j_{B}} -
  \mbox{\boldmath $r$}_{h_{S}} \mid} \right\} \nonumber \\
&& \!\!  \left. +
   {1 \over N_{S}} \!\!\int \!d\mbox{\boldmath $r$} d\mbox{\boldmath $r$}'
   \rho_{S} (\mbox{\boldmath $r$})
   n_{S} (\mbox{\boldmath $r$}')
   { [ (\mbox{\boldmath $r$}_{i_{S}} - \mbox{\boldmath $r$}) \cdot \mbox{\boldmath $v$}_{CMS}]
   \over \mid \mbox{\boldmath $r$}_{i_{S}} - \mbox{\boldmath $r$} \mid^{3} \mid \mbox{\boldmath $r$}_{j_{B}}
   - \mbox{\boldmath $r$}' \mid } \right]
\label{ESSfin}
\end{eqnarray}

\noindent
where $\alpha_{CMS} \equiv \mbox{\boldmath $x$}_{CMS} \cdot 
{(\mbox{\boldmath $r$}_{h_{S}} - \mbox{\boldmath $r$}_{j_{B}}) \over \mid \mbox{\boldmath $r$}_{h_{S}} - 
\mbox{\boldmath $r$}_{j_{B}} \mid}$.

\section{Particular limits}

\subsection{Point-mass satellites}

In eq. (\ref{EBSfin}) it is implicitly assumed that the distance between 
a generic satellite particle and a generic background particle cannot be
smaller than a scale, $d_{min}$, that is, that there exists a minimum 
effective impact parameter. As the main contribution to the integrals
appearing in this expression comes from small $\mid \mbox{\boldmath $r$}_{S} - \mbox{\boldmath $r$}_{B}
\mid$ values ($\mbox{\boldmath $r$}_{S}$ and $\mbox{\boldmath $r$}_{B}$ are generic satellite
and background particle positions), an accurate determination of the $d_{min}$
scale is a crucial point when studying dynamical friction. This scale, however,
cannot be determined by the present approach to this problem. There exist
in literature several estimates of $d_{min}$ for specific situations
(White 1976; Bontekoe \& van Albada 1987). White (1976) in fact shows that
for spherical symmetric satellites, it is a good approximation to consider them
as point-like systems with a cut-off in their distances to background
particles, and that for the case of satellites with a King mass distribution
(King 1966), this cut-off is given by $d_{min} \simeq {R_{t} \over 5}$, where
$R_{t}$ is the satellite tidal radius. A point mass satellite has
a density profile given by $\rho_{S} (\mbox{\boldmath $r$}_{S}) = M_{S} \delta(\mbox{\boldmath $r$}_{S}-
\mbox{\boldmath $r$}_{CMS})$, where $\delta(\mbox{\boldmath $r$}_{S}-\mbox{\boldmath $r$}_{CMS})$ is the delta
function in three dimensions and $\mbox{\boldmath $r$}_{CMS}$ is the position of the
satellite center-of-mass.
 With this value of $\rho_{S} (\mbox{\boldmath $r$}_{S})$,
eq. (\ref{EBSfin}) gives in the point mass limit:

\begin{eqnarray}
&&\!\!\!\!\!\!\!\!\!\!\left\langle 
\left(  {dE_{orb}^{S} \over dt} \right)_{\delta t} \right\rangle  =
\left(  {m_B \sigma_{B}^{2}
\over m_S \sigma_{S}^{2}} - 1 \right)
{G^{2}  M_{S}^{2} \over \sigma_{B}^{2}} \nonumber \\
&&\times\left[ 
\int d\mbox{\boldmath $r$}_{h_{B}} \rho_{B}(\mbox{\boldmath $r$}_{h_{B}}) 
\Theta (\mbox{\boldmath $r$}_{h_{B}})
{ (\mbox{\boldmath $r$}_{CMS} - \mbox{\boldmath $r$}_{h_{B}}) 
\cdot \mbox{\boldmath $v$}_{CMS} \over
\mid \mbox{\boldmath $r$}_{CMS} - \mbox{\boldmath $r$}_{h_{B}} \mid^{4}}
\right. \nonumber \\
&& \times \left\{ {\sigma_{B}^{2} \over \sigma_{T}^{2}}
exp(\alpha_{CMS}^{2} - x_{CMS}^{2}) \mbox{erf}(\alpha_{CMS}) 
- 1 \right\} \nonumber \\
&&+{1 \over N_{B}} \int d\mbox{\boldmath $r$} d\mbox{\boldmath $r$}'
\rho_{B} (\mbox{\boldmath $r$})
n_{B} (\mbox{\boldmath $r$}') \Theta 
 (\mbox{\boldmath $r$}) \Theta
(\mbox{\boldmath $r$}') \nonumber \\
&& \times \left. { [ (\mbox{\boldmath $r$}_{CMS} - 
\mbox{\boldmath $r$}) \cdot \mbox{\boldmath $v$}_{CMS}]
\over \mid \mbox{\boldmath $r$}_{CMS} - \mbox{\boldmath $r$} 
\mid^{3} \mid \mbox{\boldmath $r$}_{CMS}
- \mbox{\boldmath $r$'} \mid } \right] 
\label{EBSpoint}
\end{eqnarray}

\noindent
where now $\alpha_{CMS} \equiv \mbox{\boldmath $x$}_{CMS} \cdot 
{ (\mbox{\boldmath $r$}_{CMS} - \mbox{\boldmath $r$}_{h_{B}}) \over \mid \mbox{\boldmath $r$}_{CMS} 
- \mbox{\boldmath $r$}_{h_{B}} \mid}$  and 
$ \Theta ( \mbox{\boldmath $r$}) \equiv 
\Theta~(\mid~\mbox{\boldmath $r$}_{CMS} - 
\mbox{\boldmath $r$} \mid - d_{min})$ is the 
step function.

\subsection{Homogeneous backgrounds. The Chandrasekhar limit}

When the background density is uniform, we can set $\rho_{B}(\mbox{\boldmath $r$}_{B})
= \rho_{B,0}$ and $\mbox{\boldmath $r$}_{CMS} = 0$ in eq. (\ref{EBSpoint}) and then it
becomes:

\begin{eqnarray}
&&\!\!\!\!\!\!\!\!\!\!\left\langle  \left(  
{dE_{orb}^{S} \over dt} \right)_{\delta t} \right\rangle  =
\left(  {m_B \sigma_{B}^{2}
\over m_S \sigma_{S}^{2}} - 1 \right)
{ 4 \pi G^{2}  M_{S}^{2} \rho_{B,0} \ln \Lambda \over v_{CMS} }
\nonumber \\
&& \times
[ \mbox{erf}(x_{CMS}) - {2 \over \sqrt \pi } x_{CMS} \exp (- x_{CMS}^{2})].
\label{chandra}
\end{eqnarray}

This recovers the Chandrasekhar formula for the motion of a massive test
particle in an homogeneous background, except for the factor containing
the temperature ratio (which vanishes in the  Chandrasekhar formula because
$m_S \gg m_B$ in this limit), and,
 now, $\sigma_{T}=(\sigma_S^2 +\sigma_B^2)^{1/2}$ at the place of
the background velocity dispersion. 

\begin{figure}
\psfig{file=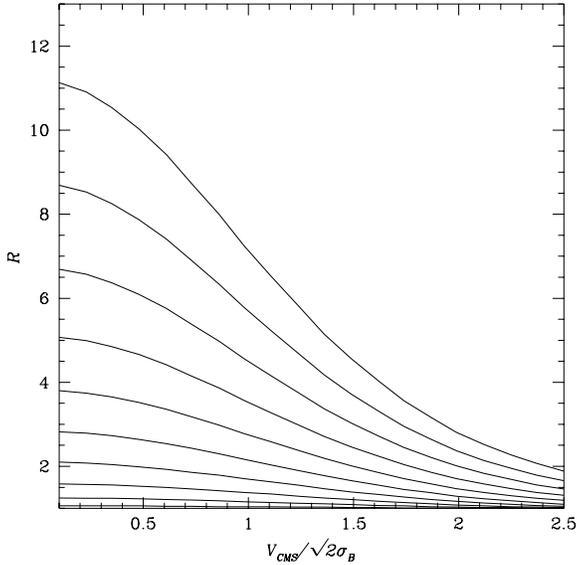,width=8cm}
\caption{Ratio, $R$, of satellite energy loss in the Chandrasekhar limit
with $\sigma_S=0$ and with $\sigma_S \neq 0$, assuming
that $T_B/T_S = 0$. Lines correspond to different $\sigma_S/\sigma_B$ 
ratios from 0.2 to 2.0 (ratio increases as the thickness of
the line increses)}
\label{ma_chan}
\end{figure}

 In the Figure \ref{ma_chan} we plot the ratio, $R$, of 
satellite energy loss in the 
Chandrasekhar limit with $\sigma_S=0$ and with $\sigma_S \neq 0$, assuming
that $T_B/T_S = 0$. As can be seen in this Figure, the effect of a non zero 
$\sigma_S$ increases with increasing $\sigma_S$ and is more important
at low $v_{CMS}$. The Chandrasekhar formula always overestimates the dynamical
friction force, and in some situations the effect of neglecting the 
satellite velocity dispersion could cause an error as high as an order
of magnitude.

\section{Summary and discussion}

We have derived an expression giving the orbital energy exchange due to
dynamical friction experienced by an extended body, composed of $N_S$ 
bound particles endowed with a velocity spectrum, as it moves interacting
with a non homogeneous discrete background. It has been assumed that both,
the satellite and the background, have Maxwellian velocity distributions and
that the background is static.

Self-interactions of both satellite and background particles have been taken
into account. This results in no effect on their energy exchange.

Heating terms appear in quite a natural way in our approach both due to 
interactions among particles of the same kind or of different kind. They
are a factor of $1/N$ smaller than the orbital effects.

In the point-like satellite limit (or small as compared with the background
size) and constant density background,
we obtain an expression that recovers, 
in the limit $\sigma_S/\sigma_B \rightarrow 0$ and $T_S/T_B \rightarrow 
\infty$, Chandrasekhar's dynamical friction formula 
(eq. (\ref{Chf})). Its comparison with the energy loss given by eq. (\ref{Chf}) 
allows for a quantification of the effects of having a non zero $\sigma_S$. It
has been found out that the energy loss is always smaller in this case, and
that the difference can be up to about an order of magnitude for slow
satellites as compared with $\sigma_B$.

In deriving eq. (\ref{dEgen}) we have considered time intervals, $\delta t$,
that are short as compared with the time scale for the variation of the particle
velocities and positions. This allows us to neglect the effects of the 
smooth gravitational potential gradients. However, in order to carry out a
precise calculation of dynamical friction, the whole history of the system
from an early enough time and the interactions along the entire satellite
trajectory should have been taken into account.
This would have made the problem extremely difficult to solve. Instead, taking
only a finite $\delta t$, means that interactions with distant particles 
have not been accurately considered. Nevertheless, we recall that the 
contribution of particles to dynamical friction quickly decreases
with distance, so that this neglect should not result in major consequences.

The bound of $\delta t$ has also another consequence: this approach
is unable to describe slowly accumulating effects on dynamical friction
(Kalnajs 1972) or the effect of reversible dynamical feedback (e.g. 
Tremaine \& Weinberg 1984), because they arise as a consequence
of the periodic motion of the satellite after many revolutions.

Despite these shortcomings, the extension of the fluctuation-dissipation
approach to dynamical friction presented in this paper, has 
resulted in the derivation of a formula that takes into account the
space and velocity structure of the satellite. This represents a common
situation in many astrophysical processes and, as we have shown, might have
important quantitative consequences in the setting-up of timescales 
for these processes.

\section*{Acknowledgments}

We thank Drs. G. Gonz\'alez-Casado and E. Salvador-Sol\'e for
interesting informations. M.A. G\'omez-Flechoso was supported by the
Direcci\'on General de Investigaci\'on Cient\'{\i}fica y T\'ecnica
(DGICYT, Spain) through a fellowship.

The DGICYT also supported in part this work , grants AEN93-0673 and
PB93-0252.

\appendix
\section{The correlation matrix}

The correlation matrix is defined in eq. (\ref{defC}) with the fluctuating
forces given by eq. (\ref{ffgen}). Each matrix is an object with four
index, and each index takes on two different values ($B$ and $S$). This
makes sixteen different possibilities, corresponding to the tensorial
product of four different possibilities for 
the $\mbox{\boldmath $F$}_{i_{\gamma}}^{\alpha}(t)$ forces. Writing the Fourier transform
of $\mid \mbox{\boldmath $r$} - \mbox{\boldmath $r$}' \mid^{-1}$ and $n_{\alpha}(\mbox{\boldmath $r$})$ with 
respect to $\mbox{\boldmath $r$}$, the convolution theorem and  eq. (\ref{ffgen})
imply that:

\begin{eqnarray}
&&\!\!\!\!\!\!\!\!\!\!\!\!
\mbox{\boldmath $F$}_{i_{\gamma}}^{\alpha}(\mbox{\boldmath $r$}_{i_{\gamma}},t) 
= - i{ G m_{\alpha} m_{\gamma} \over 2 \pi^{2}}  \int 
{d \mbox{\boldmath $k$} \over k^{2}} \mbox{\boldmath $k$}
\exp [ i \mbox{\boldmath $k$} \cdot \mbox{\boldmath $r$}_{i_{\gamma}}(t) ] 
 \nonumber \\
&& \times \left[
\sum_{g_{\alpha} \neq i_{\gamma}} \exp [ -i \mbox{\boldmath $k$} \cdot 
\mbox{\boldmath $r$}_{g_{\alpha}}(t)]
- {N'_{\alpha} \over N_{\alpha}} n_{k}^{\alpha} \right],
\label{A1}
\end{eqnarray}

\noindent
where 

\begin{equation}
n_{k}^{\alpha} \equiv \int d \mbox{\boldmath $r$} 
n_{\alpha}(\mbox{\boldmath $r$}) 
exp [- i \mbox{\boldmath $k$} \cdot \mbox{\boldmath $r$}].
\label{defnk}
\end{equation}

With this definition

\begin{equation}
\langle \exp  [- i \mbox{\boldmath $k$} \cdot \mbox{\boldmath $r$}_{i_{\alpha}} ] \rangle_{0} =
{n_{k}^{\alpha} \over N_{\alpha}}
\label{avexp}
\end{equation}

\noindent
and then the average of the stochastic forces, eq. (\ref{A1}), in the
unperturbed state vanishes, as required.

Inserting  eq. (\ref{A1}) in eq. (\ref{defC}), we get:

\begin{eqnarray}
&&\!\!\!\!\!\!\!\!\!\!\!\!C_{i_{\gamma} j_{\delta}}^{\alpha \beta} (s) =
- {G^{2} m_{\alpha} m_{\beta} m_{\gamma} m_{\delta} \over 4 \pi^{4}}
 \nonumber \\
&& \times
\int {d \mbox{\boldmath $k$} \over k^{2}} {d \mbox{\boldmath $l$} \over l^{2}} 
\mbox{\boldmath $k$} \otimes \mbox{\boldmath $l$} \exp[i(\mbox{\boldmath $k$} \cdot \mbox{\boldmath $r$}_{i_{\gamma}}(0) +
\mbox{\boldmath $l$} \cdot \mbox{\boldmath $r$}_{j_{\delta}}(s))] 
 \nonumber \\
&&\times
\left\langle \left[ 
\sum_{g_{\alpha} \neq i_{\gamma}} \exp [ -i \mbox{\boldmath $k$} \cdot
\mbox{\boldmath $r$}_{g_{\alpha}}(0)] - {N'_{\alpha} \over N_{\alpha}} 
n_{k}^{\alpha}\right] \right. \nonumber \\
&&\times \left. \left[ 
\sum_{h_{\beta} \neq j_{\delta}} \exp [ -i \mbox{\boldmath $l$} \cdot
\mbox{\boldmath $r$}_{h_{\beta}}(s)] - {N'_{\beta} \over N_{\beta}} 
n_{l}^{\beta}\right]
\right\rangle_{0}
\label{A4}
\end{eqnarray}

\noindent
and eq. (\ref{avexp}) now gives:

\begin{eqnarray}
&&\!\!\!\!\!\!\!\!\!\!\!\!C_{i_{\gamma} j_{\delta}}^{\alpha \beta} (s) =
- {G^{2} m_{\alpha} m_{\beta} m_{\gamma} m_{\delta} \over 4 \pi^{4}}
\nonumber \\
&&\times
\int {d \mbox{\boldmath $k$} \over k^{2}} {d \mbox{\boldmath $l$} \over l^{2}}
\mbox{\boldmath $k$} \otimes \mbox{\boldmath $l$} \exp[i(\mbox{\boldmath $k$} \cdot \mbox{\boldmath $r$}_{i_{\gamma}}(0) +
\mbox{\boldmath $l$} \cdot \mbox{\boldmath $r$}_{j_{\delta}}(s))] 
\nonumber \\
 &&\times
\left[ \sum_{g_{\alpha} \neq i_{\gamma}} \sum_{h_{\beta} \neq j_{\delta}}
 E_{kl} (g_{\alpha}, h_{\beta}, s) - {N'_{\alpha} \over N_{\alpha}}
 {N'_{\beta} \over N_{\beta}} n_{k}^{\alpha}  n_{l}^{\beta} \right]
\label{A5}
\end{eqnarray}

\noindent
where

\begin{equation}
 E_{kl} (g_{\alpha}, h_{\beta}, s) \equiv \langle 
 \exp \{ -i [ \mbox{\boldmath $k$} \cdot
 \mbox{\boldmath $r$}_{g_{\alpha}}(0) + 
\mbox{\boldmath $l$} \cdot
\mbox{\boldmath $r$}_{h_{\beta}}(s)] \} \rangle_{0}.
\label{A6}
\end{equation}

It is assumed that particles are uncorrelated in the unperturbed state. Then, 
if $\alpha \neq \beta$, necessarily $g_{\alpha} \neq h_{\beta}$ and
the average in eq. (\ref{A6}) factorizes, as corresponding to uncorrelated
variables, giving $C_{i_{\gamma} j_{\delta}}^{\alpha \beta} (s) =0$.
The average also factorizes for different particles belonging to the
same particle class, i.e., when $\alpha = \beta$ but 
$g_{\alpha} \neq h_{\alpha}$. There are $N'_{\alpha} 
(N'_{\alpha} -1)$
such terms, and each of them has the value ${  n_{k}^{\alpha}  n_{l}^{\alpha}
\over N_{\alpha}^{2} }$. When   $g_{\alpha} =  h_{\alpha}$, that is, 
when we consider
the same particle, the average does not factorize anymore. Taking this
considerations into account, the correlation matrix becomes:

\begin{eqnarray}
&&\!\!\!\!\!\!\!\!\!\!\!\!C_{i_{\gamma} j_{\delta}}^{\alpha \alpha} (s) =
- {G^{2} m_{\alpha}^{2} m_{\gamma} m_{\delta} \over 4 \pi^{4}}\nonumber\\
&&\times
\int {d \mbox{\boldmath $k$} \over k^{2}} {d \mbox{\boldmath $l$} \over l^{2}}
\mbox{\boldmath $k$} \otimes \mbox{\boldmath $l$} \exp[i(\mbox{\boldmath $k$} \cdot \mbox{\boldmath $r$}_{i_{\gamma}}(0) +
\mbox{\boldmath $l$} \cdot \mbox{\boldmath $r$}_{j_{\delta}}(s))]
\nonumber \\
&&\times \left[ N'_{\alpha}  E_{kl} (g_{\alpha}= h_{\alpha}, s) -  N'_{\alpha}
 { n_{k}^{\alpha}  n_{l}^{\alpha} \over  N_{\alpha}^{2}} \right].
\label{A7}
\end{eqnarray}

The position of a generic particle at time $s$ satisfying
(\ref{timeint}) can be written as:

\begin{equation}
\mbox{\boldmath $r$}_{j_{\alpha}}(s) = \mbox{\boldmath $r$}_{j_{\alpha}, 0 } +  \mbox{\boldmath $v$}_{j_{\alpha},0} s
\label{A8}
\end{equation}

An integration of eq. (\ref{eqmov}) would give four more terms corresponding
to the gravitational acceleration and the fluctuating forces. The acceleration
terms are negligible when compared with $\mbox{\boldmath $v$}_{j_{\alpha},0} s$ (see eq. (\ref{timeint})) and the fluctuating forces would give rise to third
order terms.

Once this expression for the particle positions at time $s$ is substituted
in eq. (\ref{A7}), the integrals over $d \mbox{\boldmath $k$}$ and $d \mbox{\boldmath $l$}$ can be
easily calculated taking the gradient of the Fourier representation for
the Green's function of the Laplace equation in three dimensions:

\begin{equation}
\int  {d \mbox{\boldmath $k$} \over k^{2}} \mbox{\boldmath $k$} \exp [i \mbox{\boldmath $k$} \cdot \mbox{\boldmath $A$}] =
i { 2 \pi \mbox{\boldmath $A$} \over A^{3}}.
\label{A9}
\end{equation}

Finally, combining
eqs. (\ref{defnk}),  (\ref{A6}), (\ref{A7}), (\ref{A8}) and (\ref{A9}) the
correlation matrix reads:

\begin{eqnarray}
&&\!\!\!\!\!\!\!\!\!\!\!\!C_{i_{\gamma} j_{\delta}}^{\alpha \alpha} (s) =
G^{2} m_{\alpha}^{2} m_{\gamma} m_{\delta} { N'_{\alpha} \over N_{\alpha}}
 \nonumber\\
&&\!\!\!\times 
\!\left[\! \sum_{h_{\alpha}} 
{ (\mbox{\boldmath $r$}_{i_{\gamma}} -  \mbox{\boldmath $r$}_{h_{\alpha}})
\over \mid \!\mbox{\boldmath $r$}_{i_{\gamma}} -  
\mbox{\boldmath $r$}_{h_{\alpha}}\! \mid^{3}} \otimes
{ [ \mbox{\boldmath $r$}_{j_{\delta}} -  \mbox{\boldmath $r$}_{h_{\alpha}} +
s(\mbox{\boldmath $v$}_{j_{\delta}} -  \mbox{\boldmath $v$}_{h_{\alpha}})] \over
\mid\! \mbox{\boldmath $r$}_{j_{\delta}} -  \mbox{\boldmath $r$}_{h_{\alpha}} +
s(\mbox{\boldmath $v$}_{j_{\delta}} -  \mbox{\boldmath $v$}_{h_{\alpha}}) 
\!\mid^{3}} \right. \nonumber \\
&&\!\!\! - {1 \over N_{\alpha}} \int 
d \mbox{\boldmath $r$} d \mbox{\boldmath $r$}' n_{\alpha}(\mbox{\boldmath $r$})
 n_{\alpha}(\mbox{\boldmath $r$}') \nonumber\\
&&\!\!\!\left.\times{ (\mbox{\boldmath $r$}_{i_{\gamma}} - \mbox{\boldmath $r$}) \over
 \mid \mbox{\boldmath $r$}_{i_{\gamma}} - \mbox{\boldmath $r$} 
\mid^{3}} \otimes 
 {( \mbox{\boldmath $r$}_{j_{\delta}} - \mbox{\boldmath $r$}' + s 
\mbox{\boldmath $v$}_{j_{\delta}}) \over
 \mid \mbox{\boldmath $r$}_{j_{\delta}} - \mbox{\boldmath $r$}' + s 
\mbox{\boldmath $v$}_{j_{\delta}} \mid^{3}} \right] .
\label{A10}
\end{eqnarray}

The $0$ subindex have been dropped from the $\mbox{\boldmath $r$}$ and
$\mbox{\boldmath $v$}$ vectors.

\section{}

\label{integral_I}

In this appendix we calculate the integral

\begin{equation}
\label{for1}
I = \int d \mbox{\boldmath $v$} f_M^S(\mbox{\boldmath $v$}) \exp \left[ \gamma^2 - y^2 \right] \mbox{erf}(\gamma) ,
\end{equation}

\noindent
where $\mbox{\boldmath $v$}$ is the velocity of class  $S$ particles,
that is

\begin{equation}
\label{for2}
 \mbox{\boldmath $v$} = \mbox{\boldmath $v$}_{CMS} + \mbox{\boldmath $u$} ,
\end{equation}

\noindent
whose distribution function, $f_M^S$, is a Maxwellian isotropic  
in $\mbox{\boldmath $u$}$

\begin{equation}
\label{mas}
  f_M^S(\mbox{\boldmath $v$}) = \frac{1}{(2\pi \sigma_S^2)^{3/2}} \exp \left[-\frac{\left(\mbox{\boldmath $v$}-\mbox{\boldmath $v$}_{CMS}\right)^2}{2 \sigma_S^2}\right]  .
\end{equation}

In the last equation $\mbox{\boldmath $y$} \equiv \mbox{\boldmath $v$}/\sqrt{2}\sigma_B$, $\gamma \equiv \mbox{\boldmath $y$} \cdot 
\mbox{\boldmath $\zeta$}$, 
and $\mbox{\boldmath $\zeta$}$ is a unit constant vector. Expression (\ref{for2}) 
implies:

\begin{equation}
\label{for4}
\gamma \equiv \gamma_{CMS} + \gamma_{u}
\end{equation}

\noindent
where

\[
\gamma_{CMS} = \frac{\mbox{\boldmath $v$}_{CMS}}{\sqrt{2} \sigma_B} \mbox{\boldmath $\zeta$}
\]

\noindent
 and

\[
\gamma_{u} = \frac{\mbox{\boldmath $u$}_{CMS}}{\sqrt{2} \sigma_B} \mbox{\boldmath $\zeta$}
\]

The definition of the error function and equation (\ref{for4})
imply that:

\begin{equation}
\label{for5}
\mbox{erf}(\gamma) = \mbox{erf}(\gamma_{CMS}) + \frac{2}{\sqrt{\pi}}
\int_0^{\gamma_{u}} \!dt \exp\left[-(t+\gamma_{CMS})^2\right]  .
\end{equation}

We now write $I$ (eq. (\ref{for1}))  as
an integral with respect to the relative velocities 
$\mbox{\boldmath $u$}$. Taking $\mbox{\boldmath $\zeta$} = (0,0,1)$ we obtain:

\begin{eqnarray}
\label{for6}
&&\!\!\!\!\!\!\!\!\!\!\!\!
I  = 
\frac{\exp\left(\alpha^2_{CMS} \!- \!x^2_{CMS}\right)}{(2\pi\sigma_S^2)^{3/2}}
\!\int\!\! du_x  \exp \left[-\left(Au_x\! 
+\! B_x\right)^2\right] \nonumber \\
& &\!\!\! \times \!\int\!\! du_y\exp \left[-\left(Au_y \!+ \!B_y\right)^2\right]
\!\int\!\! du_z \exp\left[-\left(\frac{u_z}{\sqrt{2}\sigma_S}\right)^2\right]
\nonumber\\
& &\!\!\!\times
\left[\mbox{erf}(\gamma_{CMS}) + \frac{2}{\sqrt{\pi}}\int_0^{\gamma_{u}} dt 
\exp\left[-\left(t + \gamma_{CMS}\right)^2 \right] \right]
\end{eqnarray}

\noindent
where $\mbox{\boldmath $x$}_{CMS} \equiv \mbox{\boldmath $v$}_{CMS}/\sqrt{2}\sigma_T$, 
$\alpha_{CMS}\equiv \mbox{\boldmath $x$}_{CMS}\cdot \mbox{\boldmath $\zeta$}$,
$\sigma_T^2 \equiv \sigma_B^2 + \sigma_S^2$, 
$A \equiv \sigma_T/\sqrt{2}\sigma_S\sigma_B$,
$B_i \equiv \frac{\sigma_S}{\sigma_B}x_{CMS, i},\, \mbox{with }i = x, y$, and 
 $\gamma_{u} = u_z/\sqrt{2}\pi \sigma_B$.
The integrals over $du_x$ and $du_y$  can be easily performed and they give:

\[
\left[ \frac{\sqrt{\pi}}{A} \mbox{erf}(\infty)\right]^2 = \frac{\pi}{A^2}
\]

The integral can now be written as:

\begin{equation}
\label{for7}
I=\frac{\sigma_H^2}{\sigma_T^2} \left[\mbox{erf}(\gamma_{CMS}) +\frac{2}{\pi}
I(a,b)\right] \exp \left[\alpha^2_{CMS}-x^2_{CMS}\right]
\end{equation}

\noindent
where

\begin{equation}
\label{i_a_b}
I(a,b) \equiv \frac{1}{a}\! \int_{-\infty}^{\infty} \!\!\!\!\! ds \, 
\exp\left[-\left(\frac{s}{a}\right)^2\right] \int_0^S\!\!dt \exp\left[-(t+b)^2
\right]
\end{equation}

\noindent
with $a \equiv \sigma_S/\sigma_H $, $b \equiv \gamma_{CMS}$,  
comes from the integration on $du_z$.

The  $I(a,b)$ integral can be evaluated as follows: first, we derive it 
with respect to $b$, the we carry out the integration with respect to $t$,
obtaining:

\begin{equation}
\label{d_i_a_b}
\frac{\partial I(a,b)}{\partial\,b} \!= \!\sqrt{\pi}\! 
\left[\frac{1}{(1+a^2)^{1/2}} \exp\left(\!-\frac{b^2}{1+a^2}\!\right) \!- 
\!\exp(-b^2)  \right]
\end{equation} 

\noindent
and, finally, an integration on $b$ leads to:

\begin{equation}
\label{i_a_b_res}
I(a,b) = I(a,0) + \frac{\pi}{2}\left[ \mbox{erf} \left[ \frac{b}{(1+a^2)^{1/2}} \right] -\mbox{erf}(b) \right]
\end{equation}

\noindent
with  $ I(a,0) = 0$ because  $\mbox{erf}(s)= -\mbox{erf}(-s)$.
	Substituting the expression for $I(a,b)$ in eq. 
(\ref{for7}), we get:

\begin{equation}
\label{i_res}
I = \frac{\sigma_B^2}{\sigma_T^2}\exp\left(\alpha_{CMS}^2 - x_{CMS}^2 \right) \mbox{erf}(\alpha_{CMS})
\end{equation}

\noindent
where now  $\mbox{\boldmath $x$}_{CMS} \equiv 
\mbox{\boldmath $v$}_{CMS}/\sqrt{2}\sigma_T$ and 
$\alpha_{CMS} \equiv \mbox{\boldmath $x$}_{CMS}\cdot \mbox{\boldmath $\zeta$}$.

\label{lastpage}

\begin{thebibliography}{400}
\bibitem{Beke92} Bekenstein, J.~D. \& Maoz, E., 1992, ApJ, 390, 79
\bibitem{Beke91} Bekenstein, J.~D. \& Zamir, D., 1991, ApJ,  359, 427
\bibitem{Binn87} Binney, J. \& Tremaine, S., 1987, Galactic Dynamics.  
   Princeton Univ. Press, Princeton
\bibitem{Bont87} Bontekoe, T.~R. \& van Albada, T.~S., 1987,  MNRAS, 224, 349
\bibitem{Chan43} Chandrasekhar, S., 1943,  ApJ, 97, 255
\bibitem{Cohe75} Cohen, L., 1975, in Hayli, A., ed., Proc. IAU Symp. 60:
   Dynamics of Stellar Systems. Reidel, Dordrecht, p.~33
\bibitem{Kaln72} Kalnajs, A.~J., 1972, in Lecar, M., ed.,
   Proc. IAU Colloq. 10: Gravitational N-body Problem. Reidel, Dordrecht, p.~13
\bibitem{Kand80} Kandrup, H.~E., 1980, Phys. Rep., 63, 1
\bibitem{King66} King, I., 1966, AJ, 71, 64
\bibitem{Kubo59} Kubo, R., 1959, in Lectures in Theoretical Physics, Univ.
   of Colorado, vol. 1. Intersciences Publishers, p. 120
\bibitem{LinT83} Lin, D. N.~C. \& Tremaine, S., 1983, ApJ, 264, 364
\bibitem{Lynd72} Lynden-Bell, D. \& Kalnajs, A., 1972, MNRAS, 157, 1
\bibitem{Mamo93} Mamon, G.~A., 1993, in Combes, F.  \& Athanassoula, E., eds., 
   Gravitational Dynamics and the N-body Problem. Meudon, Paris, p. 188
\bibitem{Maoz93} Maoz, E., 1993, MNRAS, 263, 75
\bibitem{Maro68} Marochnik, L. S., 1968, SvA, 11, 873
\bibitem{Palm85} Palmer, P.~L. \& Papaloizou, J., 1985, MNRAS,  215, 691
\bibitem{Reif65} Reif, F., 1965, Fundamentals of Statistical and 
   Thermal Physics.  McGraw-Hill, New York
\bibitem{Rose57} Rosenblunth, M. N., Mc Donald, W. M. \& Judd, D. L., 1957,
   Phys. Rev., 107, 1
\bibitem{Sasl85} Saslaw, W. C., 1985, Gravitational Physics of Stellar 
   and Galactic Systems. Cambridge Univ. Press, Cambridge
\bibitem{Trem84} Tremaine, S. \& Weinberg, M.~D., 1984, MNRAS, 209, 729
\bibitem{Wein86} Weinberg, M.~D., 1986, ApJ, 300, 93
\bibitem{Wein89} Weinberg, M.~D., 1989, MNRAS, 239, 549
\bibitem{Whit76} White, S. D.~M., 1976, MNRAS, 174, 467
\bibitem{Whit83} White, S. D.~M., 1983, ApJ, 274, 53
\end{thebibliography}
\end{document}